\documentclass[final,1p,times]{elsarticle}

\usepackage{amssymb}
\usepackage{amsmath}

\usepackage{natbib}
\biboptions{numbers,sort&compress}
\usepackage[colorlinks=true,linkcolor=blue,citecolor=blue,urlcolor=blue]{hyperref}
\usepackage{subfigure}
\usepackage{graphicx}
\usepackage{latexsym}
\usepackage{bm} 
\usepackage{amsfonts}
\usepackage{soul,color,xcolor} 
\usepackage{stfloats} 
\usepackage{booktabs} 
\usepackage{tabularx}
\usepackage{tabu}
\usepackage{multirow}
\usepackage{url}
\usepackage{geometry}

\usepackage{threeparttable} 
\usepackage{makecell} 
\usepackage{bbding} 
\usepackage{algorithm}
\usepackage{algorithmic}

\graphicspath{{Figures/}} 

\journal{Elsevier}

\begin{document}
	
	\begin{frontmatter}
		
		
		
		
		
		\title{Closed-loop domain generation enhances battery health diagnosis}
		
		
		\author{Yue Wu$^{a}$, 
			Ziyu Mo$^{a}$,
			Jinze Dai$^{a}$,
			Ziqing Xia$^{a}$,
			Shengyu Tao$^{b}$,
			Yusheng Zheng$^{c}$,
			Yunhong Che$^{d,e *}$
		}
		

		\address{
			{$^{a}$ School of Electronic Information, Central South University, Changsha, 410075, China \\}
			{$^{b}$ Department of Electrical Engineering, Chalmers University of Technology, Gothenburg, 41296, Sweden\\}
			{$^{c}$ Department of Energy Technology, Aalborg University, Aalborg 9220, Denmark\\}
			{$^{d}$ College of Mechanical and Vehicle Engineering, Chongqing University, Chongqing, 400044, China\\}
			{$^{e}$ Department of Chemical Engineering, Massachusetts Institute of Technology, Cambridge, MA 02139, USA\\}
			
			{$^{*}$ Corresponding author: Yunhong Che (yunhche@mit.edu)\\}
		}
		
		\begin{abstract}
			Reliable electrode-level degradation diagnosis is essential for accurate health estimation and non-destructive monitoring in advanced battery management. Diagnosis models developed under controlled laboratory conditions can fail when operating conditions change, while acquiring sufficient full-lifespan data in new target domains remains prohibitively time-consuming and practically unfeasible. Here, we propose a closed-loop domain generation framework to enable cross-domain battery health diagnosis under dynamic conditions. Physics-constrained domain generation is achieved through Variational Autoencoder-based latent extrapolation, which maps the source data and few-shot early-life target domain data into a shared latent space and generates cross-domain samples within physical boundaries. Bayesian optimization closes the generation-diagnosis loop by automatically calibrating the sampling, latent regularization, extrapolation, perturbation, and generation-volume hyperparameters using leakage-free development validation, thereby strictly excluding later-life target data from optimization and ensuring the quality and quantity of generated data. Numerical results demonstrate that, across four cross-domain cases, the proposed framework achieves average root mean square errors of 1.53\%, 2.15\%, and 2.02\% for positive electrode active material loss, negative electrode active material loss, and lithium inventory loss, respectively, whereas traditional methods suffer from notably higher errors or even complete failure. This framework provides a general pathway for onboard and physics-consistent electrode-level degradation diagnosis, independent of downstream diagnosis models.
		\end{abstract}
		
		
	\end{frontmatter}
	
	
	\section{Introduction}
	
	Accurate battery health diagnosis is paramount for the safety, reliability, and lifespan optimization of lithium-ion batteries in electric vehicles. During long-term operation, batteries undergo complex mechanical and electrochemical degradation, leading to continuous capacity fade and increased internal resistance \cite{edge2021lithium, o2022lithium}. Developing robust degradation models and diagnostic frameworks is therefore a core function of advanced battery management \cite{reniers2019review}.
	
	In recent years, data-driven approaches have achieved remarkable success in macro-level battery capacity estimation/prediction. Unlike model- or mechanism-based methods that require complex parameterization, data-driven approaches directly capture nonlinear degradation patterns from easily measurable signals \cite{song2026deep}. Severson et al. \cite{severson2019data} pioneered a machine learning pipeline to accurately predict the cycle life using solely early-cycle data before significant capacity degradation occurs. Zhu et al. \cite{zhu2022data} developed a data-driven capacity estimation method by extracting features from voltage relaxation curves. However, these methods primarily treat the battery as a black box and focus on estimating overall capacity fade using carefully designed features. In addition, macro-capacity is merely a symptom of aging, which obscures the internal degradation mechanisms.
	
	To overcome the limitations of macro-level capacity estimation, battery degradation diagnosis is now shifting towards degradation mode analysis at the electrode level. Loss of lithium inventory and loss of active material in both electrodes can be quantified by reconstructing the full-cell open-circuit voltage (OCV) from half-cell potentials \cite{birkl2017degradation}. Based on this principle, OCV curves can be reconstructed \cite{zhu2023lithium} through scaling and translation of electrode potentials, through which the aging mode can be determined. In real-world scenarios where vehicles often undergo incomplete charging, these electrode aging parameters can also be estimated accurately using partial charging profiles \cite{tian2021electrode}. Electrochemical impedance spectroscopy and deep learning are incorporated to improve degradation diagnosis under limited labeled data \cite{sun2025deep}. More recently, material-specific diagnostic models have been developed to distinguish the degradation of silicon and graphite within composite anodes \cite{wan2026managing}. These studies demonstrate the value of mapping measurable electrical signals to physically interpretable states of battery degradation at the electrode level.
	
	Despite these advances, most diagnostic methods still rely heavily on standardized discharging profiles collected in laboratory environments and require long test time, such as C/40 \cite{che2026battery}. In practical applications, battery degradation is affected by dynamic currents, partial operating windows, temperature variations, and measurement noise. More importantly, realistic operating profiles may alter the degradation trajectory. Geslin et al. \cite{geslin2025dynamic} showed that dynamic discharge profiles produced substantially different lifetime behavior from conventional constant-current cycling. To improve diagnosis under realistic conditions, researchers have combined field data with OCV and impedance models \cite{li2022battery}, transformed high-rate measurements into low-rate-equivalent diagnostic features \cite{che2026battery}, and introduced residual learning to correct model errors under unseen operating conditions \cite{che2026mechanistically}. Other approaches use random partial operational data \cite{che2025diagnostic}, fuse arbitrary charging segments with relaxation features \cite{ding2026decoding}, and incorporate physics-informed learning \cite{tao2025non} for onboard degradation diagnosis.
	
	The difference between laboratory and field data represents a typical Lab2Real domain shift \cite{liu2025multi}. Moreover, cross-domain problems may arise from differences in operating profiles, current rates, temperatures, battery chemistries, and manufacturers. Collecting full-lifespan labeled data for every new target domain is costly and time-consuming, meaning that only limited early-life data are usually available. Existing studies have addressed this few-shot challenge through reference-cell selection and model updating \cite{che2022data}, domain adaptation without extensive target domain labels \cite{lu2023deep}, and physics-guided prediction from minimal experiments \cite{zhang2026discovery}.  Meta-learning and domain-generalization strategies are used to identify transferable early-cycle features and improve lifetime prediction under unseen operating conditions \cite{zhang2026meta}. Generative learning has also been used to expand pulse-response data for state-of-health and remaining-capacity estimation under random retirement conditions \cite{tao2024generative,tao2025immediate}. However, these studies mainly focus on macroscopic health indicators rather than electrode-level degradation modes.
	
	One promising approach to addressing the few-shot challenge is domain generation. Examples ranging from unsupervised cross-domain image generation \cite{taigman2016unsupervised} to simulation-based real-world robot training \cite{tobin2017domain} have demonstrated its potential to bridge distribution-shift gaps. For battery health diagnosis, however, the goal is not merely to reproduce signal distributions, but to preserve physically meaningful electrode-level state of health (eSOH) and aging mode evolution. Several challenges therefore still remain: (1) eSOH is governed by coupled changes in electrode capacities, lithium inventory, and utilization windows. Generated samples must preserve these relationships and remain consistent with the corresponding full-cell OCV response; otherwise, statistically reasonable data may represent unrealistic degradation states. (2) Full-lifespan eSOH labels are generally unavailable in new operating conditions, while only limited early-life target-domain data can usually be obtained, making later-life aging mode evolution difficult to infer. (3) Domain generation must bridge the electrical-feature distribution gap while providing sufficient coverage of feasible aging trajectories and preserving the mapping between dynamic operating features and degradation states. How to jointly control physical consistency, target-domain relevance, and generation quantity therefore remains an open problem.
	
	To address the aforementioned gaps, this paper proposes a novel domain generation framework for robust cross-domain battery health diagnosis. Unlike existing studies that focus on advanced diagnostic methods or on extracting new effective features, we focus on improving data quality to enhance diagnostic performance, a focus that can be extended to any diagnostic method. Three distinct contributions make this paper different from existing studies:
	
	\begin{enumerate}[(1)]
		\item A unified closed-loop domain generation framework is proposed to address the severe cross-domain shift in battery health diagnosis.
		
		\item A physics-constrained domain generation mechanism is developed by coupling variational autoencoder (VAE)-based latent extrapolation with automated Bayesian optimization (BO). The generation boundaries and generated data volume are automatically calibrated under leakage-free validation, avoiding unconstrained data generation.
		
		\item Extensive cross-domain evaluations across multiple cross-domain scenarios demonstrate the superiority of the proposed framework. 
	\end{enumerate}
	
	The remainder of this paper is organized as follows. Section \ref{model} introduces the preliminaries, including the battery half-cell model, feature extraction, and different cross-domain scenarios. Section \ref{method} details the proposed closed-loop domain generation framework. Section \ref{result} presents a comprehensive evaluation of the generated domains, diagnosis performance enhancements, and ablation studies. Finally, the main conclusions are drawn in Section \ref{conclusion}.
	
	\section{Physics-informed data foundation and domain definition \label{model}}
	
	\subsection{Battery aging mechanisms and half-cell OCV model}
	
	To enable non-destructive electrode aging diagnosis, full-cell OCV models based on half-cell reconstruction have been proposed \cite{bloom2005differential} and used for identification of loss of lithium inventory (LLI), loss of active material at the positive electrode ($\text{LAM}_{pe}$), and loss of active material at the negative electrode ($\text{LAM}_{ne}$) \cite{dubarry2012synthesize, han2014comparative, lee2020electrode}. The full-cell OCV is mathematically formulated by matching the positive and negative half-cell potentials:
	
	\begin{equation}
		\label{eq:ocv}
		\text{OCV}(z) = \text{OCP}_{p}(y) - \text{OCP}_{n}(x),
	\end{equation}
	where $z$ is the depth of discharge (DOD), defined as one minus the state of charge (SOC). $\text{OCP}_{p}$ and $\text{OCP}_{n}$ are the half-cell open-circuit potentials of the cathode and anode, respectively. $y$ and $x$ are the corresponding electrode lithiation states. The full cell charge state is related to the half-cell lithiation states:
	
	\begin{equation}
		z = \frac{Q}{C} = \frac{y-y_{100}}{y_{0}-y_{100}} = \frac{x_{100}-x}{x_{100}-x_{0}},
	\end{equation}
	where $Q$ and $C$ are the available and total capacity of the full cell. $y_{100}$, $x_{100}$, $y_{0}$, and $x_{0}$ indicate the voltage utilization window. Subscript 100 for maximum voltage, 0 for minimum voltage. The full cell capacity can be defined by the individual electrode capacities:
	
	\begin{equation}
		C = C_{p}(y_{0}-y_{100}) = C_{n}(x_{100}-x_{0}),
	\end{equation}
	where $C_{p}$ and $C_{n}$ are capacities of the positive and negative electrodes, respectively. By eliminating $C$ in the above equations, the normalized lithiation states at each electrode are:
	
	\begin{equation}
		y = y_{100} + \frac{Q}{C_{p}}, \quad  x = x_{100} - \frac{Q}{C_{n}}.
	\end{equation}
	
	Then, the full-cell OCV is parameterized with the electrode parameters [$C_{p}$, $C_{n}$, $y_{100}$, $x_{100}$]:
	
	\begin{equation}
		\label{eq:ocv1}
		\text{OCV}(z) = \text{OCP}_{p}(y_{100} + \frac{Q}{C_{p}}) - \text{OCP}_{n}(x_{100} - \frac{Q}{C_{n}}).
	\end{equation}
	
	In this study, electrode parameters [$C_{p}$, $C_{n}$, $y_{100}$, $x_{100}$] for the 92 battery cells across the whole lifespan \cite{geslin2025dynamic} are optimized by differential evolution algorithm, see details in \cite{che2026battery}. Note that the cathode exhibits minimal shift across cycling, $y_{100}$ is set to zero to avoid multi-solution problems, i.e., the fitting results are similar \cite{che2026battery}. 
	
	Once the electrode parameters are determined, LLI, $\text{LAM}_{pe}$, and $\text{LAM}_{ne}$ can be quantified \cite{lee2020electrode}:
	
	\begin{equation}
		\begin{cases}
			\label{eq:qloss}
			\text{LLI} = 1 - n_{Li}^a/n_{Li}^f, \\
			\text{LAM}_{pe} = 1 - C_{p}^a/C_{p}^f, \\
			\text{LAM}_{ne} = 1 - C_{n}^a/C_{n}^f, 
		\end{cases}
	\end{equation}
	where $n_{Li}$ is the amount of intercalated lithium in both electrodes and can be calculated by $n_{Li} = \frac{3600}{F}(y_{100}C_{p} + x_{100}C_{n})$, superscript $a$ and $f$ represent aged and fresh.
	
\begin{figure}[!h]
		\centering
		\includegraphics[width=8cm]{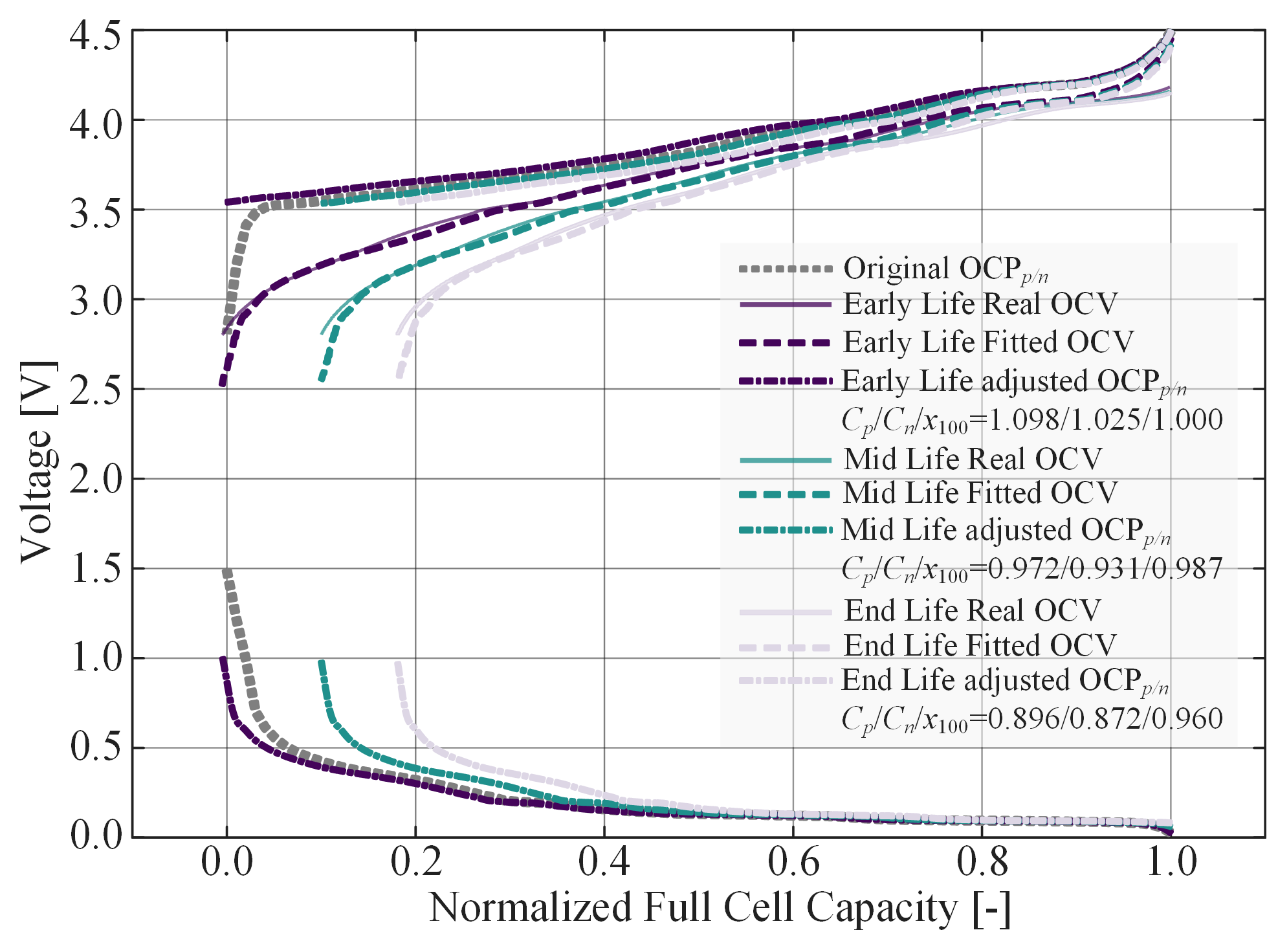}
		\caption{Illustration of the half-cell OCV model, taking the early/middle/end life of cell 77 as examples.} 
		\label{fig1_halfcell}
	\end{figure}

	An example of the half-cell OCV model is provided in Fig. \ref{fig1_halfcell}, where the early, middle, and end life OCV curves of cell 77 are plotted with the corresponding fitted [$C_{p}$, $C_{n}$, $x_{100}$] values.
	
	\subsection{Diagnosis feature extraction under dynamic discharging profiles}
	
\begin{figure}[!h]
		\centering
		\includegraphics[width=8cm]{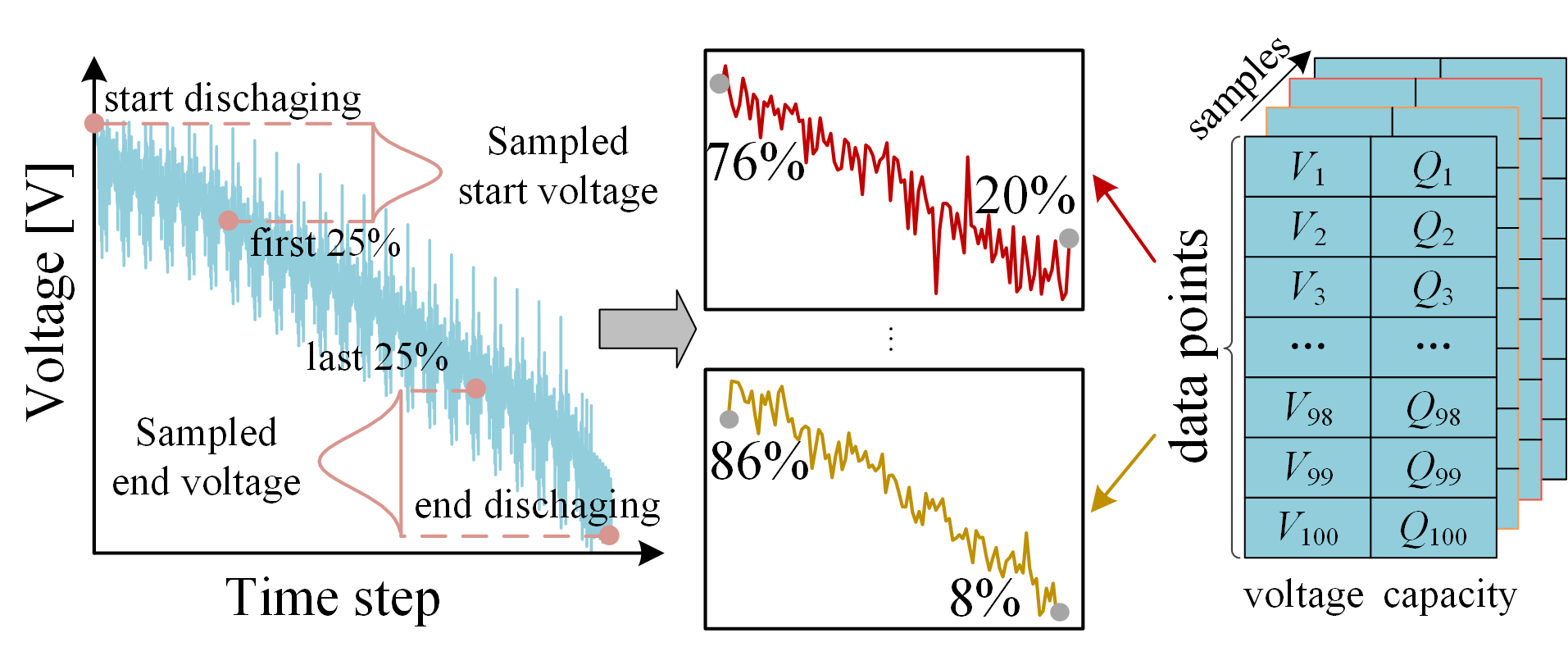}
		\caption{Random discharging feature extraction.}
		\label{fig2_feature}
	\end{figure}

	To extract reliable aging features from highly irregular dynamic discharging conditions, this study adopts the feature extraction method from random discharge segments. Specifically, from the dynamic discharging segments randomly sampled during onboard operations, voltage ($V$), local cumulative capacity ($Q$), and equivalent full cycle (EFC) are extracted to capture the battery's thermodynamic and polarization states under diverse discharging conditions. Then, all extracted dynamic feature sequences are standardized to a fixed length of 100 data points using a resampling algorithm to meet the input tensor dimensions of subsequent data-driven models, as shown in Fig. \ref{fig2_feature}. 
	
	
	\subsection{Dataset description and cross-domain definition}
	
	This study utilizes the dynamic aging dataset published by Stanford
	University \cite{geslin2025dynamic}. The dataset includes 92 commercial lithium-ion cells subjected to meticulously designed discharge profiles, ranging from conventional constant current (CC) and periodic pulses to highly dynamic synthetic and real-world electric vehicle driving cycles. From the perspective of average discharging rates, C/16, C/10, C/5, and C/2 are included to represent different operation conditions. All 92 cells share the same charging protocols. Detailed cell discharge information can be found in Table S4 of \cite{geslin2025dynamic}.
	
	\begin{figure}[!b]
		\centering
		\includegraphics[width=8cm]{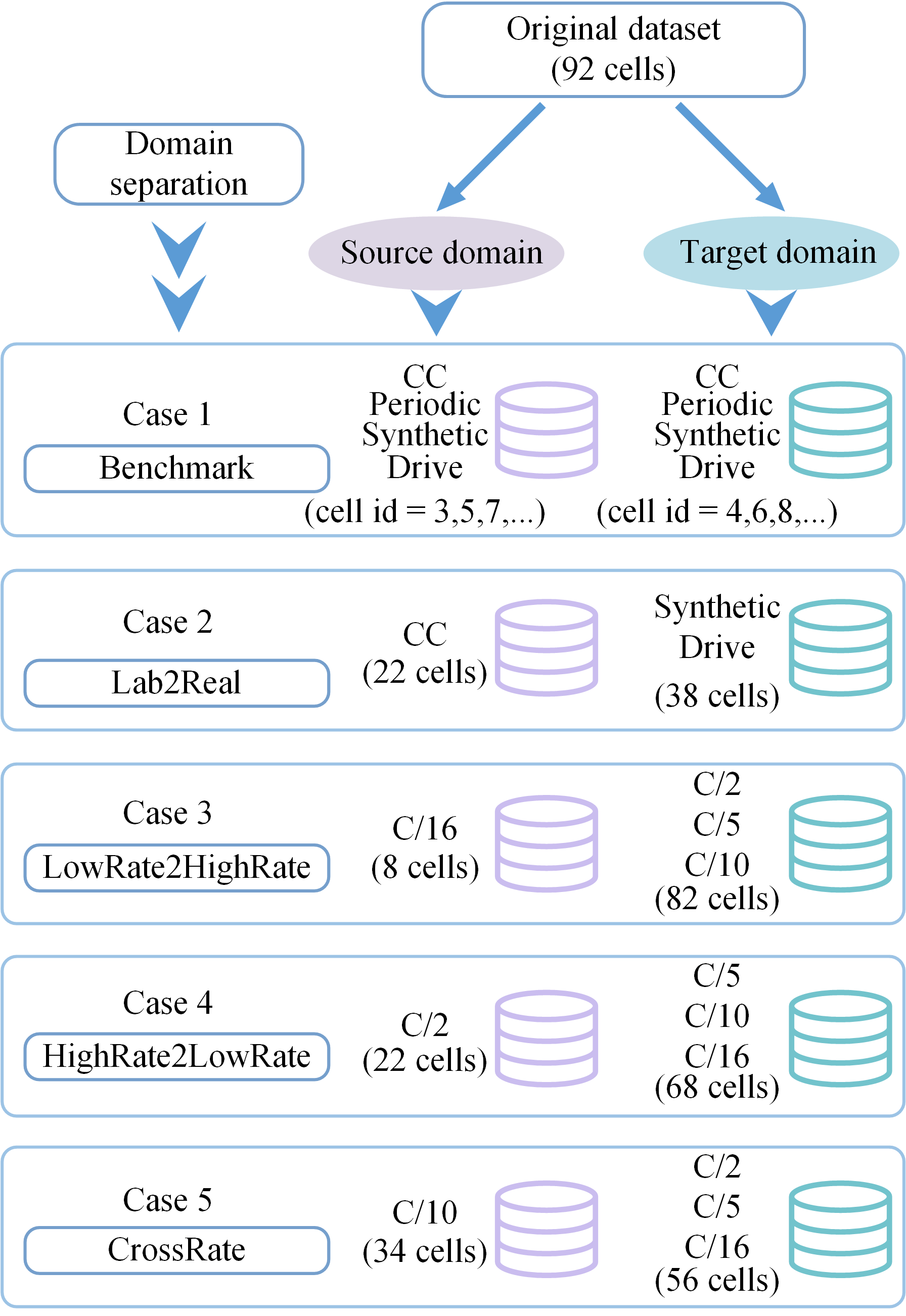}
		\caption{Different domain-separation cases.}
		\label{fig3_domain}
	\end{figure}
	
	We design five distinct cross-domain scenarios that reflect practical challenges in battery management systems, as shown in Fig. \ref{fig3_domain}. In \textbf{Case 1 Benchmark}, the source domain (training set) and the target domain (testing set) are different but share the same discharge conditions, which represents the baseline performance achievable with a sufficiently large cross-domain dataset. \textbf{Case 2 Lab2Real} indicates that only CC standard laboratory profiles are available, whereas health diagnosis must be performed under dynamic driving conditions. \textbf{Case 3 LowRate2HighRate} and \textbf{Case 4 HighRate2LowRate} evaluate generalizability across different load intensities, addressing scenarios in which historical data from low usage is used to diagnose aggressive aging, or vice versa. Note that Case 3 represents an extreme situation of data scarcity, where the training set is very limited (only 8 cells). Finally, \textbf{Case 5 CrossRate} challenges the health diagnosis by extrapolating health states across a broad spectrum of unseen current rates from a single C/10 baseline. These defined scenarios explicitly represent different levels of domain shift in actual applications. Note that Cell 17 and 84 are excluded from all five cases since Cell 18 and 83 with the same discharging protocol are failed \cite{geslin2025dynamic}, thus 90 cells are used.
	
	\section{Battery health diagnosis based on closed-loop domain generation\label{method}}
	
	To address the challenge of cross-domain battery health diagnosis with few-shot target domain data, this section details the proposed closed-loop domain generation framework, in which physics-constrained domain generation is implemented through VAE-based latent extrapolation and automated Bayesian optimization.
	
\begin{figure*}[!h]
		\centering
		\includegraphics[width=16cm]{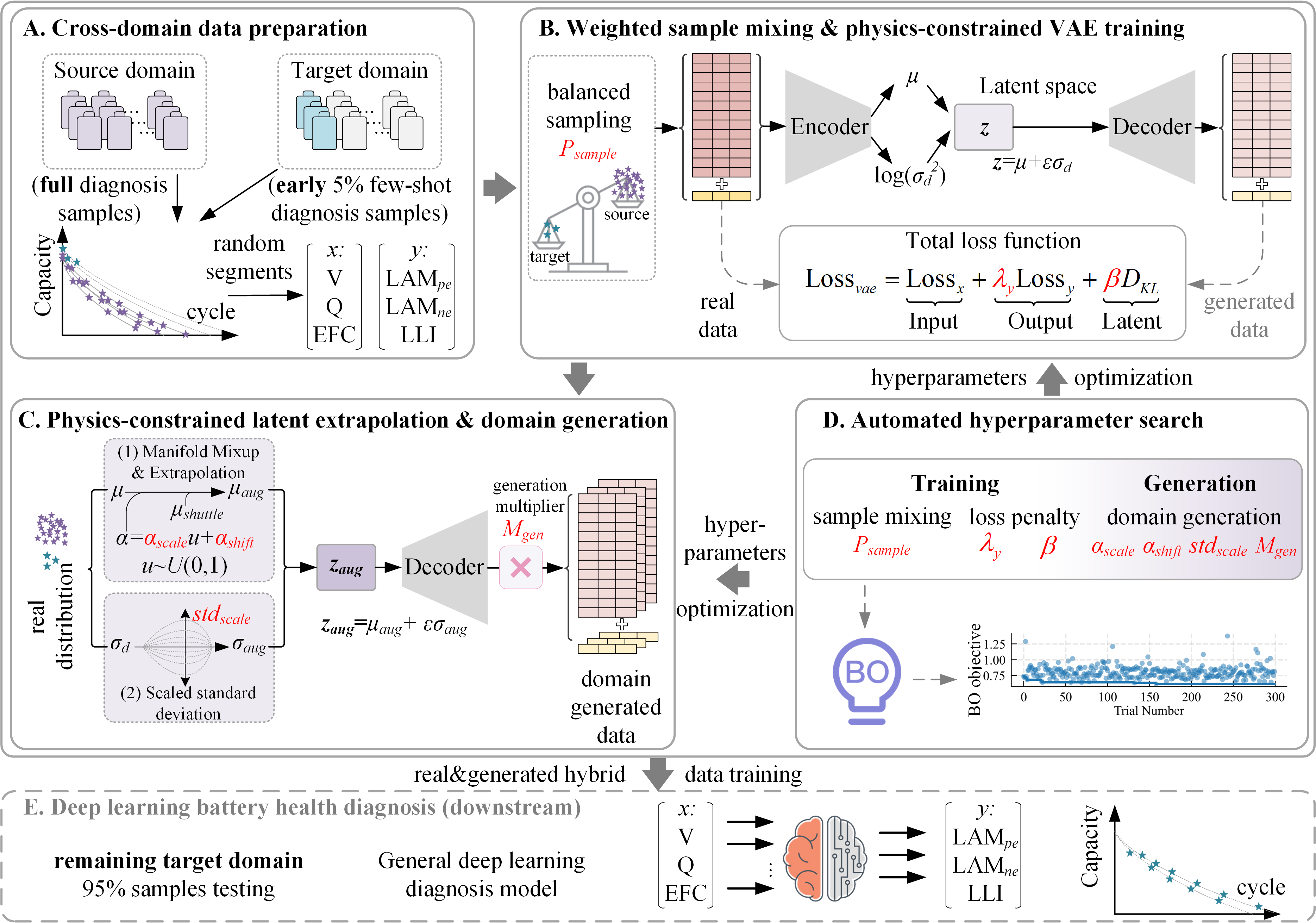}
		\caption{Closed-loop domain generation enhances battery health diagnosis.} 
		\label{fig4_structure}
	\end{figure*}

	As illustrated in Fig. \ref{fig4_structure}, the source and target domains are first separated according to the predefined cross-domain scenario, while only early-life samples of the target domain are used as few-shot information, and the remaining target domain data are strictly isolated for final testing. The available source and early-life target samples are subsequently balanced and encoded into a shared latent space using a VAE, in which dynamic electrical features and electrode-level degradation labels are jointly reconstructed. Latent-space mixup, controlled extrapolation, and stochastic perturbation are then applied to generate synthetic cross-domain samples. The generation process is automatically calibrated through Bayesian optimization using leakage-free development validation, thereby jointly constraining lifespan extrapolation and target domain adaptation. Finally, the optimized generated samples are combined with the available real data to train a general downstream diagnostic model, which estimates $\text{LAM}_{pe}$, $\text{LAM}_{ne}$, and $\text{LLI}$ and is evaluated on the remaining unseen target domain lifespan.
	
	\subsection{Early few-shot sampling of the target domain}
	
	In practical electric vehicle operations, acquiring full-lifecycle degradation data is generally unfeasible. Consequently, diagnosing battery health utilizing only early-life data, e.g., the first 5\% of the lifespan, represents a realistic scenario. To establish a rigorous few-shot evaluation protocol, a specific sampling mechanism is formulated. For a given battery $i$ in the target domain comprising $N_i$ total degradation sequences, the number of sampled sequences $N_{sample, i}$ is determined by Eq. (\ref{eq:sample_n}):
	\begin{equation}
		N_{sample, i} = \max(1, \lfloor N_i \times 0.05 \rfloor),
		\label{eq:sample_n}
	\end{equation}
	where $\lfloor \rfloor$ means round-down operation.
	
	These initial $N_{sample, i}$ sequences are extracted to form the few-shot target domain dataset. Subsequently, the indices of these sampled sequences are globally marked. The remaining $N_i - N_{sample, i}$ sequences are strictly isolated to form the unseen test set, thereby preventing data leakage during model evaluation.
	
	\subsection{Physics-constrained domain generation via VAE-based latent extrapolation}
	
	\subsubsection{Cross-domain data representation and balanced sampling}
	
	The domain generation module employs a Variational Autoencoder (VAE) \cite{kingma2013auto} to map the source domain data and the few-shot target domain data into a shared latent space. The input vector $v_{in} \in \mathbb{R}^{303}$ is constructed by concatenating the flattened dynamic feature vector $v_{x} \in \mathbb{R}^{300}$ (reshaped from the raw matrix $x \in \mathbb{R}^{100 \times 3}$) and the diagnosis label vector $y \in \mathbb{R}^3$, i.e., $\text{LAM}_{pe}$, $\text{LAM}_{ne}$, and $\text{LLI}$. 
	
	To balance the learning process between the source and target domains, a specific sampling probability $P_{sample}$ is defined. The sampling weights for individual source domain samples $w_{src}$ and target domain samples $w_{tgt}$ are calculated below:
	\begin{equation}
		w_{src} = \frac{1 - P_{sample}}{N_{src}},
		\label{eq:weight_src}
	\end{equation}
	\begin{equation}
		w_{tgt} = \frac{P_{sample}}{N_{tgt}},
		\label{eq:weight_tgt}
	\end{equation}
	where $N_{src}$ and $N_{tgt}$ denote the total number of samples in the source and target domains, respectively.
	
	\subsubsection{Physics-constrained joint loss reconstruction}
	
	The VAE loss function integrates the reconstruction losses and the Kullback-Leibler (KL) divergence. To ensure the generated features adhere to the physical degradation states, a weight parameter $\lambda_y$ is applied to the label reconstruction loss. The loss $\text{Loss}_{vae}$ is expressed in Eq. (\ref{eq:vae_loss}):
	\begin{equation}
		\text{Loss}_{vae} = \text{Loss}_{x} + \lambda_y \text{Loss}_{y} + \beta D_{KL},
		\label{eq:vae_loss}
	\end{equation}
	where $\text{Loss}_{x}$ and $\text{Loss}_{y}$ represent the Mean Squared Error (MSE) for the feature and label reconstructions, defined below:
	\begin{equation}
		\text{Loss}_{x} = \frac{1}{M} \sum_{j=1}^{M} (x_j - \hat{x}_j)^2,
		\label{eq:lx}
	\end{equation}
	\begin{equation}
		\text{Loss}_{y} = \frac{1}{K} \sum_{k=1}^{K} (y_k - \hat{y}_k)^2,
		\label{eq:ly}
	\end{equation}
	where $M = 300$ and $K = 3$ denote the dimensions of the flattened dynamic features and the degradation indicators, respectively. Furthermore, $D_{KL}$ is calculated based on the mean $\mu$ and variance $\sigma^2$ of the latent distribution \cite{kingma2013auto}:
	\begin{equation}
		D_{KL} = -\frac{1}{2} \sum_{d=1}^{D} \left(1 + \log(\sigma_d^2) - \mu_d^2 - \sigma_d^2\right).
		\label{eq:kld}
	\end{equation}
	
	\subsubsection{Latent space extrapolation and data decoding}
	
	During the generation phase, an extrapolation mechanism is introduced. The scaling factor $\alpha$ controls the extrapolation range and is dynamically generated utilizing a standard uniform random variable $u \sim \mathcal{U}(0, 1)$, an extrapolation scaling coefficient $\alpha_{scale}$, and an extrapolation shift parameter $\alpha_{shift}$, as defined in Eq. (\ref{eq:alpha}):
	\begin{equation}
		\alpha = u \cdot \alpha_{scale} + \alpha_{shift}.
		\label{eq:alpha}
	\end{equation}
	
	The augmented mean $\mu_{aug}$ is subsequently computed by interpolating the original mean $\mu$ with a randomly shuffled mean sequence $\mu_{shuffle}$, i.e., the Manifold Mixup \cite{verma2019manifold}, formulated in Eq. (\ref{eq:mu_aug}):
	\begin{equation}
		\mu_{aug} = \alpha \mu + (1 - \alpha) \mu_{shuffle}.
		\label{eq:mu_aug}
	\end{equation}
	
	The final latent vector $z_{aug}$ is obtained via the reparameterization trick \cite{kingma2013auto, rezende2014stochastic} with a standard deviation scaling coefficient $std_{scale}$, as shown in Eq. (\ref{eq:z_aug}):
	\begin{equation}
		z_{aug} = \mu_{aug} + \epsilon \cdot (std_{scale} \sigma),
		\label{eq:z_aug}
	\end{equation}
	where $\epsilon \sim \mathcal{N}(0, I)$. The detailed training and generation procedure is summarized in Algorithm \ref{alg:vae}.
	
	\begin{algorithm}[!h]
		\caption{VAE-based latent extrapolation for physics-constrained domain generation}
		\label{alg:vae}
		\begin{algorithmic}[1]
			\REQUIRE Source dataset $D_{src}$, target few-shot dataset $D_{tgt}$, hyperparameters $P_{sample}$, $\lambda_y$, $\beta$, $\alpha_{scale}$, $\alpha_{shift}$, $std_{scale}$, $M_{gen}$
			\ENSURE Generated synthetic dataset $D_{gen}$
			\STATE Initialize VAE network parameters
			\STATE Calculate sample weights $w_{src}$ and $w_{tgt}$ using Eq. (\ref{eq:weight_src}) and Eq. (\ref{eq:weight_tgt})
			\STATE Construct dataloader with weighted random sampler
			\FOR{epoch = 1 to 500}
			\FOR{each batch $[X, Y]$ in dataloader}
			\STATE Compute $\mu, \sigma^2$ via encoder
			\STATE Sample latent variable $z$ via reparameterization
			\STATE Compute reconstructions $\hat{x}, \hat{y}$ via decoder
			\STATE Compute total loss $\text{Loss}_{vae}$ using Eq. (\ref{eq:vae_loss})
			\STATE Update network parameters via backpropagation
			\ENDFOR
			\ENDFOR
			\STATE Encode concatenated dataset $D_{src} \cup D_{tgt}$ to obtain $\mu, \sigma^2$
			\STATE Generate $\mu_{shuffle}$ by randomly permuting $\mu$
			\STATE Initialize empty synthetic dataset $D_{gen} \leftarrow \emptyset$
			\FOR{$m = 1$ to $M_{gen}$}
			\STATE Calculate scaling factor $\alpha$ using Eq. (\ref{eq:alpha})
			\STATE Calculate $\mu_{aug}$ and $z_{aug}$ using Eq. (\ref{eq:mu_aug}) and Eq. (\ref{eq:z_aug})
			\STATE Decode $z_{aug}$ to obtain current synthetic batch and append to $D_{gen}$
			\ENDFOR
			\RETURN $D_{gen}$
		\end{algorithmic}
	\end{algorithm}
	
	\subsection{Automated hyperparameter search via Bayesian optimization}
	
	To systematically identify an effective hyperparameter configuration without accessing the unseen target domain data, Bayesian optimization \cite{bergstra2011algorithms,attia2020closed} is embedded into the closed-loop domain generation framework. The optimization process aims to determine the hyperparameter set $\boldsymbol{\phi}^*$ that minimizes a leakage-free objective $J(\boldsymbol{\phi})$, as formulated in Eq. (\ref{eq:bo_obj}):
	
	\begin{equation}
		\boldsymbol{\phi}^* = \arg\min_{\boldsymbol{\phi} \in \mathcal{H}} J(\boldsymbol{\phi}),
		\label{eq:bo_obj}
	\end{equation}
	where $\mathcal{H}$ defines the predefined search space.
	
\begin{figure}[!h]
		\centering
		\includegraphics[width=6.5cm]{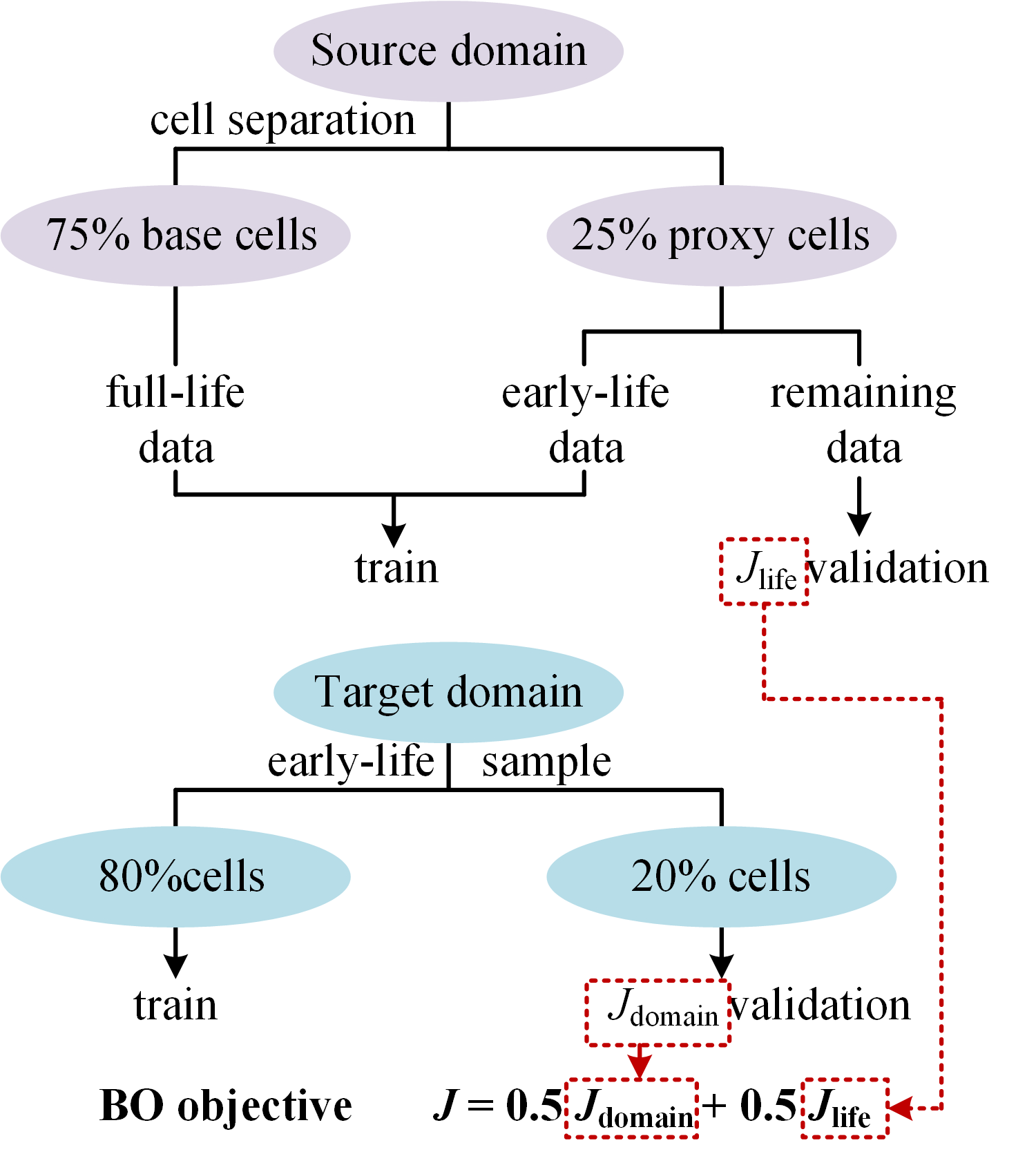}
		\caption{Leakage-free validation for Bayesian hyperparameter optimization.}
		\label{fig_bo_validation}
	\end{figure}

	Since the remaining 95\% of the target domain data must be strictly isolated for final evaluation, it cannot be directly used to assess the full-lifespan extrapolation capability during hyperparameter optimization. Therefore, a lifespan-aware surrogate validation protocol is constructed using the fully available source domain data, together with an independent validation of the visible early-life target domain data, as illustrated in Fig. \ref{fig_bo_validation}.
	
	Specifically, the source domain cells are randomly divided at the cell level into 75\% source-base cells and 25\% source-proxy cells. The complete data of the source-base cells are used for training, whereas only the first 5\% of the data of the source-proxy cells are used during training, and the remaining 95\% are reserved for validation. The target domain cells are further divided into 80\% training cells and 20\% validation cells, whose visible early-life samples are used for training and target domain validation, respectively. All remaining 95\% of the data from each target domain cell is excluded from the complete BO procedure and retained solely for the final independent test.
	
	For each candidate hyperparameter set, the VAE generation pipeline and downstream diagnosis model are trained using the available source and target training subsets. To prevent cells containing different numbers of diagnostic fragments from dominating the optimization, the root mean square error (RMSE) is first calculated independently for each validation cell and each degradation indicator. For validation subset $s$, the cell-level RMSE is defined as:
	
	\begin{equation}
		\mathrm{RMSE}_{i,k}^{(s)}
		=
		\sqrt{
			\frac{1}{N_{i,s}}
			\sum_{j=1}^{N_{i,s}}
			\left(
			\hat{y}_{i,j,k}-y_{i,j,k}
			\right)^2
		},
		\label{eq:bo_cell_rmse}
	\end{equation}
	where $i$ denotes the battery cell, $k \in \{\mathrm{LAM}_{pe},\mathrm{LAM}_{ne},\mathrm{LLI}\}$ denotes the degradation indicator, and $N_{i,s}$ is the number of validation samples for cell $i$.
	
	The RMSE values are subsequently averaged across cells and degradation indicators. The lifespan-surrogate validation objective $J_{\mathrm{life}}$ and the target domain validation objective $J_{\mathrm{domain}}$ are expressed as:
	
	\begin{equation}
		J_s =
		\frac{1}{3|\mathcal{C}_s|}
		\sum_{i \in \mathcal{C}_s}
		\sum_{k}
		\mathrm{RMSE}_{i,k}^{(s)},
		\qquad
		s \in \{\mathrm{domain},\mathrm{life}\},
		\label{eq:bo_subobjective}
	\end{equation}
	where $\mathcal{C}_s$ denotes the set of cells contained in the corresponding validation subset.
	
	The final BO objective equally considers the capability to extrapolate degradation trajectories over the lifespan and the capability to adapt to the target domain distribution:
	
	\begin{equation}
		J(\boldsymbol{\phi})
		=
		0.5J_{\mathrm{domain}}
		+
		0.5J_{\mathrm{life}}.
		\label{eq:bo_fx}
	\end{equation}
	
	This validation protocol provides informative feedback for hyperparameter selection while ensuring that no late-life target domain label participates in model training, validation, or hyperparameter optimization.
	
	The search space $\mathcal{H}$ comprises seven explicit hyperparameters $\boldsymbol{\phi} = [P_{sample}, \lambda_y, \beta, \alpha_{scale}, \alpha_{shift}, std_{scale}, M_{gen}]$, which can be classified into three aspects:
	
	\textbf{(1) Sample mixing control}: The target domain sampling probability $P_{sample}$ determines the proportion of few-shot target data integrated into each training batch.
	
	\textbf{(2) Loss penalty terms}: The label reconstruction weight $\lambda_y$ balances the waveform fidelity and degradation accuracy, while the KL divergence weight $\beta$ controls the continuity of the latent space.
	
	\textbf{(3) Domain generation control}: The extrapolation coefficients $\alpha_{scale}$ and $\alpha_{shift}$ determine the bounds of the feature mixup, the noise scaling factor $std_{scale}$ governs the micro-perturbation diversity, and the generation multiplier $M_{gen}$ explicitly defines the overall volume of the synthetic samples generated from the latent space.
	
	To navigate this search space efficiently, the Tree-structured Parzen Estimator algorithm \cite{bergstra2011algorithms} is employed. By constructing probabilistic models of the objective function, the algorithm iteratively selects hyperparameters that maximize the expected improvement, eventually converging to an optimized cross-domain generation.
	
	
	\subsection{Downstream battery health diagnosis model based on deep learning}
	
	The proposed closed-loop domain generation framework is decoupled from downstream diagnosis network architectures. In this study, a Multilayer Perceptron (MLP) is adopted as the baseline model to evaluate the effectiveness of the generated data. Note that any other networks such as long short-term memory, convolutional neural network, Transformer, and other architectures can also be used here.
	
	The MLP acts as a mapping function $f_\theta: \mathbb{R}^{300} \rightarrow \mathbb{R}^3$, transforming the flattened dynamic feature vector $x \in \mathbb{R}^{300}$ into the diagnosis vector $y \in \mathbb{R}^3$. To satisfy the physical requirement that degradation metrics remain non-negative, a Softplus activation function is applied to the final output layer. The Softplus function is provided below:
	\begin{equation}
		\text{Softplus}(\cdot) = \log(1 + \exp(\cdot)).
		\label{eq:softplus}
	\end{equation}
	
	During model training, an unweighted MSE is employed to maintain the stability of gradient calculation. The diagnosis loss $\text{Loss}_{diag}$ is defined in Eq. (\ref{eq:mlp_loss}):
	\begin{equation}
		\text{Loss}_{diag} = \text{MSE}_{\text{LAM}_{pe}} + \text{MSE}_{\text{LAM}_{ne}} + \text{MSE}_{\text{LLI}}.
		\label{eq:mlp_loss}
	\end{equation}
	
	The model is trained on the combined dataset consisting of the source domain samples, the early-life few-shot target domain samples, and the generated synthetic samples. Subsequently, its performance is evaluated exclusively on the isolated remaining target domain test set.
	
	\section{Results and discussion \label{result}}
	
	This section evaluates the closed-loop domain generation framework for battery health diagnosis. First, the limitations of the purely data-driven baseline model are analyzed under out-of-distribution (OOD) scenarios. Then, the generated feature distributions and hyperparameter optimization processes are validated. Finally, the enhancement in diagnostic performance and the effects of various data sampling strategies are systematically evaluated across multiple cross-domain cases. Diagnostic performance is quantified using RMSE, mean absolute error (MAE), and the coefficient of determination ($R^2$). Note that Case 2 will be the example presented in detail, as it represents the most common Lab2Real cross-domain scenario. Other cases are provided in the Supplementary Materials.
	
	\subsection{Baseline diagnosis failure in out-of-distribution scenarios}
	
	To establish a benchmark, the baseline method is evaluated under Case 2, in which the model is trained solely on the source domain (CC discharging profiles) and tested on the unseen target domain (dynamic synthetic and real-world driving profiles).
	
\begin{figure*}[!h]
		\centering
		\includegraphics[width=16cm]{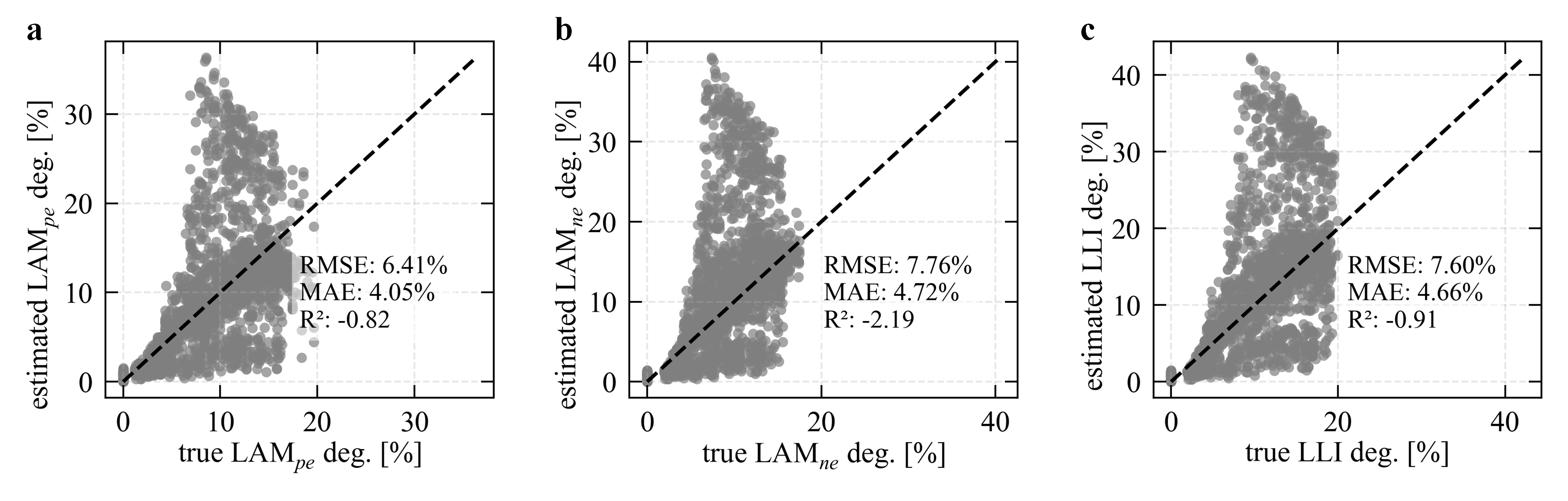}
		\caption{Battery health diagnosis failure using the baseline method under Case 2. (a) for LAM$_{pe}$, (b) for LAM$_{ne}$, (c) for LLI. All CC discharging samples are used for training, with all Synthetic/Drive discharging samples for testing.} 
		\label{fig6_baseline_failure}
	\end{figure*}

	The baseline model exhibits significant estimation errors in cross-domain testing, with RMSE values for $\text{LAM}_{pe}$, $\text{LAM}_{ne}$, and $\text{LLI}$ reaching 6.41\%, 7.76\%, and 7.60\%, respectively. The R$^2$ values are entirely negative. Fig. \ref{fig6_baseline_failure} further illustrates the diagnostic failure of the baseline method when confronting OOD dynamic features. The model exhibits severe overestimation and large estimation variance. Notably, a large number of estimated points deviate upward from the reference line, with estimation error exceeding 30\% or 40\% while the true degradation remains below 20\%. This extreme dispersion demonstrates that purely data-driven neural networks fail to establish robust feature-to-label mappings under severe domain shifts, thereby completely losing estimation reliability in the absence of prior target domain guidance.
	
	\subsection{Visualization and evaluation of physics-constrained domain generation}
	
	To address the baseline method's failure to estimate under severe domain shifts, the physics-constrained domain generation approach is introduced to bridge the feature-distribution gap between the source and target domains. To optimize this generation process, Bayesian optimization is executed for 300 trials to systematically tune 7 critical hyperparameters. 
	
	The search spaces are carefully bounded based on physical and algorithmic intuitions. Specifically, the target sampling probability $P_{sample}$ is searched within $[0.0, 1.0]$ to explore the optimal domain mixing ratio. For the loss penalty terms, the label reconstruction weight $\lambda_y \in [1.0, 50.0]$ promotes consistency with the physically interpretable degradation labels, while the log-scaled KL divergence weight $\beta \in [0.001, 0.5]$ preserves dynamic feature details without causing posterior collapse. In terms of latent exploration, the extrapolation coefficients $\alpha_{scale} \in [1.0, 3.0]$ and $\alpha_{shift} \in [-1.0, 0.0]$ are restricted to ensure safe feature expansion while limiting latent exploration to physically feasible regions. Finally, the noise scaling factor $std_{scale} \in [0.5, 3.0]$ and the generation multiplier $M_{gen} \in \{1, 2, \dots, 10\}$ are constrained to provide adequate perturbation diversity and synthetic data volume without causing noise domination.
	
	\begin{figure*}[!h]
		\centering
		\includegraphics[width=16cm]{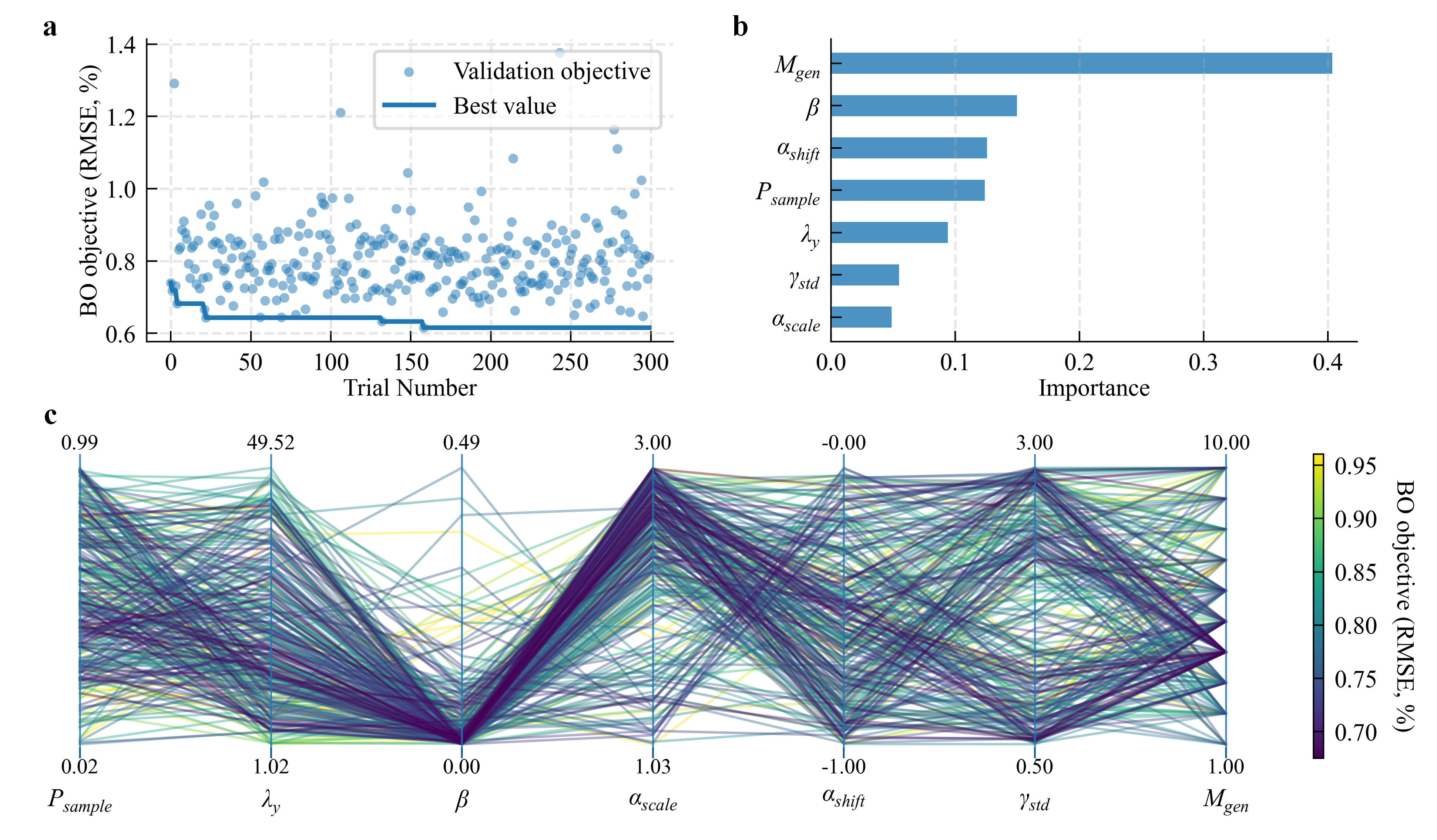}
		\caption{Hyperparameter search of domain generation using Bayesian optimization for Case 2. (a) Objective value of all 300 trials. (b) Quantitative evaluation of the importance of the seven hyperparameters. (c) Parallel-coordinate representation of the seven hyperparameters and objective values.} 
		\label{fig7_hyperparameter}
	\end{figure*}
	
	Fig. \ref{fig7_hyperparameter}(a) illustrates the convergence behavior of the initial 300-trial Bayesian optimization. The best development objective decreases rapidly during the early search stage and is further improved in a stepwise manner as the search proceeds, eventually reaching a stable plateau. This indicates that the optimizer progressively narrows the feasible hyperparameter region rather than relying on a single early optimum.
	
	The importance analysis in Fig. \ref{fig7_hyperparameter}(b) shows that the generation multiplier $M_{gen}$ is the most influential hyperparameter in Case 2, with an importance substantially higher than other parameters. The KL-divergence weight $\beta$ exhibits the second-highest importance, followed by the extrapolation shift $\alpha_{shift}$ and the target domain sampling probability $P_{sample}$. In contrast, $\alpha_{scale}$ contributes comparatively little to the variance of the development objective within the predefined search interval. The parallel coordinate results in Fig. \ref{fig7_hyperparameter}(c) further demonstrate that trials with similar objective values may correspond to completely different combinations of individual parameters, confirming the strongly coupled nature of the domain generation process.
	
	The dominant influence of $M_{gen}$ indicates that, for the Lab2Real transition in Case 2, the amount of generated data is a major factor governing the balance between enriching the feature distribution and preserving the information contained in the available real samples. A high importance score does not imply that a larger $M_{gen}$ is always preferable; rather, it indicates that inappropriate generation volumes cause significant changes in downstream diagnostic performance. The relatively high importance of $\beta$ further suggests that controlling latent space regularization remains important when mapping regular CC degradation features and highly dynamic driving features into a shared representation. Meanwhile, the effects of $\alpha_{shift}$ and $P_{sample}$ show that both the location of latent space exploration and the relative contribution of the few-shot target samples remain relevant to the generation process.
	
	The comparatively low importance of $\alpha_{scale}$ in the present Case 2 results differs from the intuitive expectation that the extrapolation range alone should dominate a severe Lab2Real shift. This result suggests that, once the extrapolation range is restricted to a physically reasonable interval, diagnostic performance becomes more sensitive to how many samples are generated and how the latent distribution and domain mixture are regularized. Therefore, the optimized generation process is governed by the interaction among data volume, latent space regularization, domain sampling, and extrapolation, rather than by a single extrapolation parameter.
	
\begin{figure*}[!h]
		\centering
		\includegraphics[width=16cm]{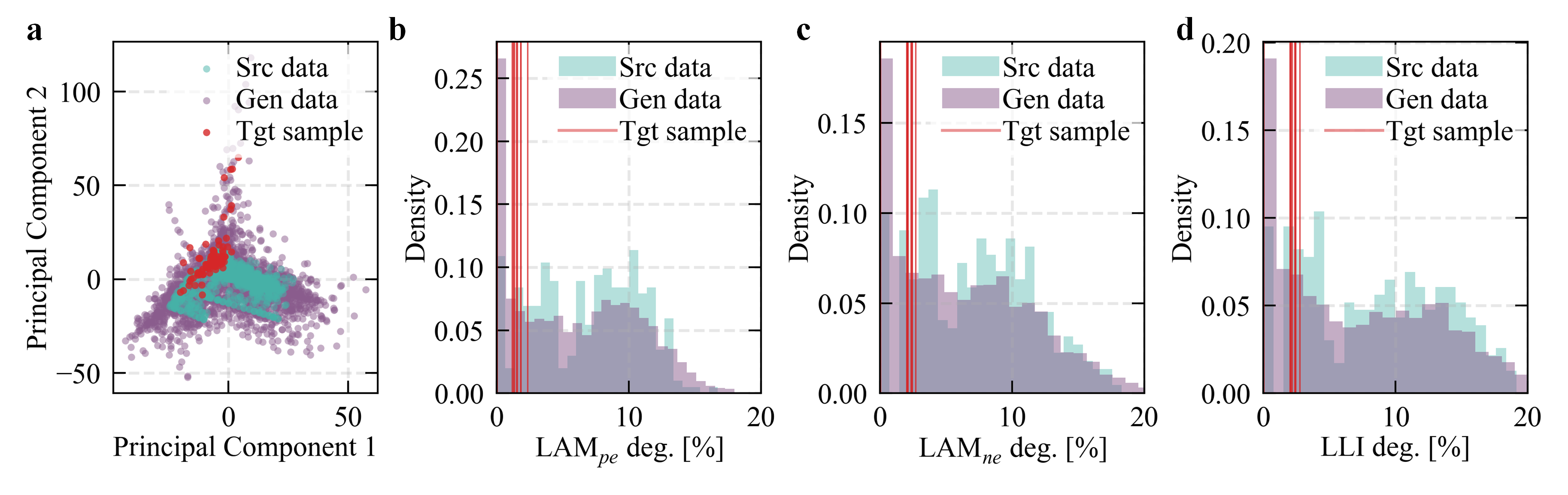}
		\caption{Visualization of domain generation for Case 2. (a) The PCA scatter results of the VAE latent vector. (b), (c), and (d) are the LAM$_{pe}$, LAM$_{ne}$, and LLI distributions of the source domain data, sampled target domain data, and generation domain data. 'Src' refers to the source, 'Tgt' to the target, and 'Gen' to the generation.} 
		\label{fig8_domaingeneration}
	\end{figure*}

	Using the optimized hyperparameters, Fig. \ref{fig8_domaingeneration} visualizes the data distributions to validate the effectiveness of physics-constrained domain generation. As shown by the principal component analysis (PCA) of the latent space in Fig. \ref{fig8_domaingeneration}(a), a distinct distribution gap is evident between the source domain data and the sampled few-shot target domain data, particularly in the upper-left region, confirming the severe domain shift in the Lab2Real scenario. After domain generation, the generated samples extend from the source distribution toward the target-oriented region and populate the intermediate latent space, thereby improving the continuity of feature coverage without producing visually apparent unrestricted extrapolation. This indicates that the jointly optimized sampling, latent regularization, extrapolation, and generation parameters effectively reshape the source distribution toward the target domain. Furthermore, the degradation histograms in Fig. \ref{fig8_domaingeneration}(b)-(d) qualitatively compare the distributions of the source domain samples, few-shot target domain samples, and generated samples. Since the target data are restricted to the first 5\% of the lifespan, their degradation states are naturally concentrated within a narrow early-life range. In contrast, the generated samples retain the broader degradation coverage learned from the source domain while enriching the sample density toward the target distribution. Consequently, the generated domain provides a more continuous training distribution in both the latent feature space and the degradation label space.
	
	\subsection{Diagnostic performance enhancement driven by physics-constrained domain generation}
	
\begin{figure*}[!h]
		\centering
		\includegraphics[width=16cm]{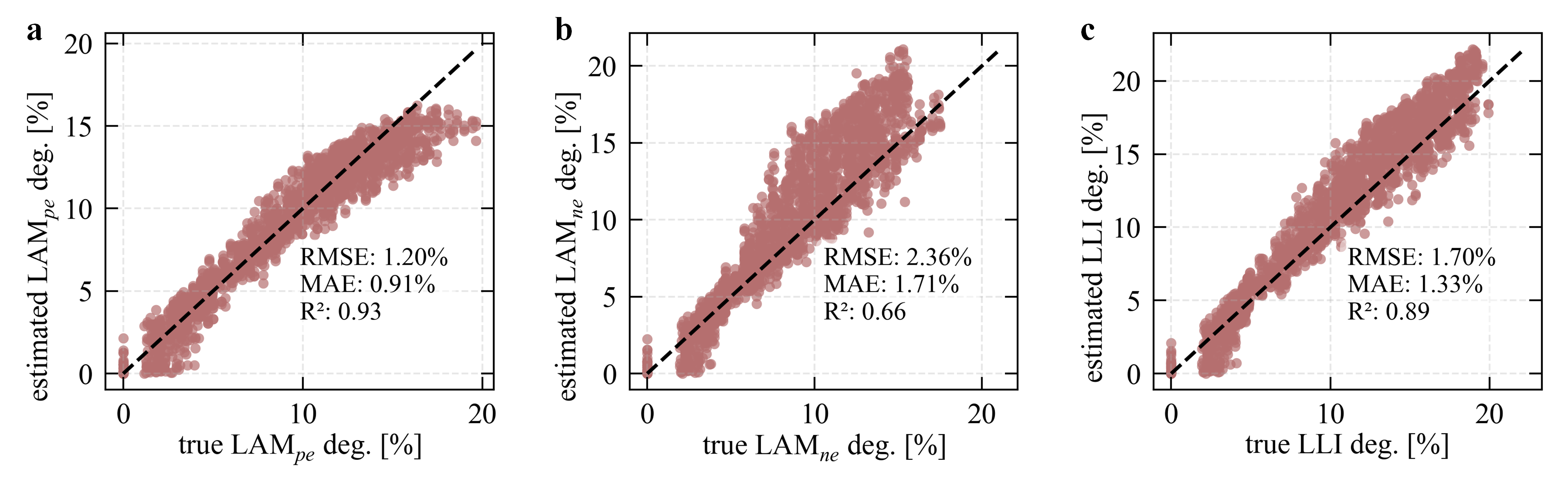}
		\caption{Battery health diagnosis results of the proposed domain generation method under Case 2. (a) for LAM$_{pe}$, (b) for LAM$_{ne}$, (c) for LLI. All CC discharging samples and the first 5\% few-shot Synthetic/Drive samples are used for training, with the remaining Synthetic/Drive samples for testing.} 
		\label{fig9_vae_enhancement}
	\end{figure*}

	To evaluate the improvement in estimation achieved by the proposed physics-constrained domain generation framework, the diagnostic performance under Case 2 is first analyzed. As presented in Fig. \ref{fig9_vae_enhancement}, the proposed method significantly reduces the estimation errors, achieving RMSE values of 1.20\%, 2.36\%, and 1.70\% for $\text{LAM}_{pe}$, $\text{LAM}_{ne}$, and $\text{LLI}$, respectively. Meanwhile, the R$^2$ values for $\text{LAM}_{pe}$, $\text{LAM}_{ne}$, and $\text{LLI}$ are improved to 0.93, 0.66, and 0.89, respectively. Compared with Fig. \ref{fig6_baseline_failure}, the estimated scatter points are more closely aligned with the diagonal reference line. Compared with the baseline method, the proposed method effectively mitigates severe overestimation and exhibits a more constrained error distribution. This demonstrates that the physics-constrained domain generation process produces data that enable the diagnostic model to establish robust mappings, yielding enhanced generalization to the unseen target domain with dynamic cycling.
	
\begin{figure*}[!h]
		\centering
		\includegraphics[width=16cm]{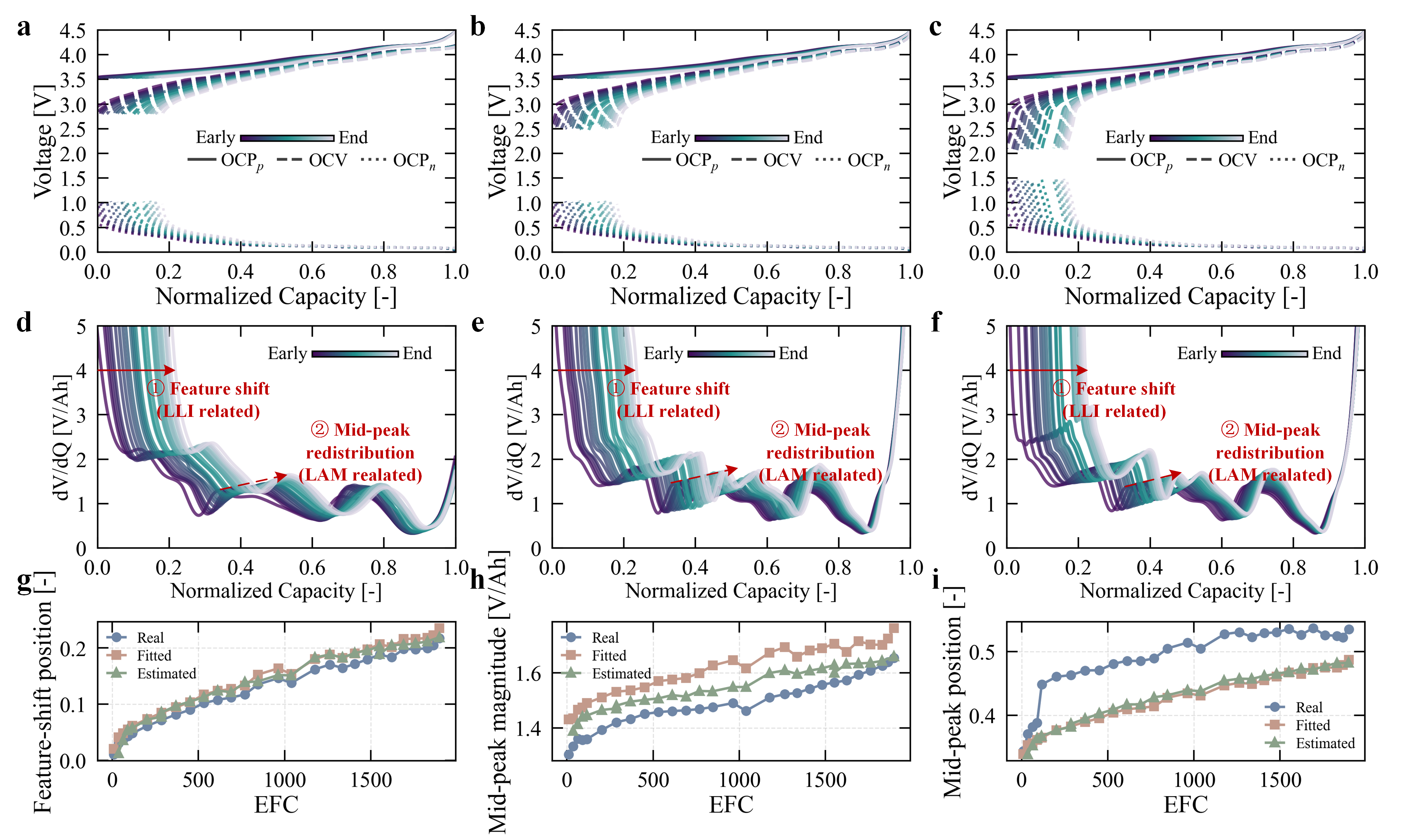}
		\caption{Battery half-cell OCV model degradation results under Case 2, taking cell 77 as an example. (a) real half-cell OCV curves, (b) fitted half-cell OCV curves, (c) estimated half-cell OCV curves by the proposed method, (d) real dV/dQ curves, (e) fitted dV/dQ curves, and (f) estimated dV/dQ curves by the proposed method, (g) comparison of high-gradient feature shift position changes, (h) comparison of mid-peak magnitude changes, and (i) comparison of mid-peak position changes.} 
		\label{fig10_cell77}
	\end{figure*}

	Furthermore, the physical consistency of the diagnosed degradation modes is examined through half-cell OCV reconstruction and differential-voltage analysis, as shown in Fig. \ref{fig10_cell77}. Taking cell 77 under Case 2 as an example, Figs. \ref{fig10_cell77}(a)--(c) compare the real, fitted (from differential evolution), and estimated evolutions of OCP$_p$, full-cell OCV, and OCP$_n$ throughout aging. The three groups exhibit highly consistent electrode-level evolution. In particular, OCP$_p$ remains relatively stable over most of the capacity range, whereas the accessible OCP$_n$ window and the corresponding low-voltage region of the full-cell OCV gradually shift with aging. This behavior is consistent with the coupled evolution of electrode parameters: LAM$_{pe}$ and LAM$_{ne}$ modify the effective electrode capacities through $C_p$ and $C_n$, while LLI alters the relative utilization window through $x_{100}$. Aging is reflected as a redistribution and displacement of the accessible electrode reaction windows rather than a simple uniform voltage offset. 
	
	We further examine the physical consistency using the more sensitive dV/dQ curves in Figs. \ref{fig10_cell77}(d)--(f). Two representative aging features are highlighted. First, the high-gradient feature marked by \textcircled{1} shifts along the normalized capacity axis. Using the position at $dV/dQ=4$ V/Ah as a quantitative indicator, Fig. \ref{fig10_cell77}(g) shows an increase from approximately 0.01--0.02 at early life to about 0.21--0.23 near the end of life. Both the fitted and estimated trajectories closely reproduce this monotonic shift, reflecting the change in electrode utilization windows and thus is sensitive to LLI-related degradation. Second, the characteristic mid-capacity peak marked by \textcircled{2} exhibits simultaneous evolution in magnitude and position. As shown in Fig. \ref{fig10_cell77}(h), the mid-peak magnitude increases continuously with EFC, and the estimated trajectory closely follows the fitted reference over the whole aging process. Meanwhile, Fig. \ref{fig10_cell77}(i) shows that the peak position moves from approximately 0.35 toward 0.5 as aging proceeds, reflecting redistribution of the electrode reaction windows associated with LAM.
	
	Note that the small residual difference between the real and fitted feature positions in Fig. \ref{fig10_cell77}(i) does not directly represent a diagnosis error of the proposed method. This difference mainly originates from the differential evolution-based half-cell OCV fitting used to generate the reference degradation labels. In contrast, the close agreement between the fitted and estimated trajectories confirms that the proposed method accurately recovers these physically interpretable degradation states.
	
	\subsection{Ablation study on data composition and domain generation}
	
	To systematically evaluate the individual contributions of different data compositions and domain generation, a comprehensive ablation study is conducted based on the configurations defined in Table \ref{tab1}. The comparative methods evaluate the absence of generation in the Baseline and Method 1, the absence of the first 5\% of few-shot target domain data in Method 2, the absence of source domain data in Method 3, and an idealized global sampling scenario in Method 4 against the proposed method.
	
	The quantitative results under Case 2, presented in Table \ref{tab2}, reveal the specific impact of each data composition. Incorporating only the first 5\% of target domain data via Method 1 slightly reduces the baseline errors, but the RMSE remains very high. For instance, the RMSE for $\text{LAM}_{pe}$ reaches 4.67\%, indicating that simple early-life data fine-tuning is insufficient to overcome severe domain shifts. Conversely, using samples generated solely from the source domain in Method 2 significantly improves performance, highlighting the intrinsic power of the VAE-based latent generation module. Completely discarding the source domain in Method 3 causes immediate model collapse, confirming that source domain data is necessary to model fundamental physical degradation pattern. 
	
	\begin{table*}[!h]
		\caption{Comparison of training data compositions and generation strategies of different methods.}
		\label{tab1}
		\centering
		\begin{threeparttable}
			\resizebox{16cm}{!}{
				\begin{tabular}{llll}
					\toprule
					Method & Training dataset & Testing dataset & Domain generation\\
					\midrule
					Baseline & Source domain samples & All target domain samples & \XSolidBrush\\
					Method 1 & Source domain samples + First 5\% target domain samples & Remaining 95\% target domain samples & \XSolidBrush\\
					Method 2 & Source domain samples + Generation samples & All target domain samples & \Checkmark\\
					Method 3 & First 5\% target domain samples + Generation samples & Remaining 95\% target domain samples & \Checkmark\\
					Method 4 & Source domain samples + Entire life random 5\% target domain samples + Generation samples & Remaining 95\% target domain samples & \Checkmark (data leakage)\\
					Proposed & Source domain samples + First 5\% target domain samples + Generation samples & Remaining 95\% target domain samples & \Checkmark\\
					\bottomrule
			\end{tabular}}
		\end{threeparttable}
	\end{table*}
	
	\begin{table*}[!h]
		\caption{RMSE/MAE/R$^{2}$ results of LAM$_{pe}$, LAM$_{ne}$, and LLI using different diagnosis methods under different cross-domain cases. For each case, the results of the best method (excluding Method 4) are shown in bold.}
		\label{tab2}
		\centering
		\begin{threeparttable}
			\resizebox{16cm}{!}{
				\begin{tabular}{llcccccc}
					\toprule
					Case & Index & Baseline & Method 1 & Method 2 & Method 3 & Method 4 & Proposed \\
					\midrule
					
					& RMSE/MAE/R$^{2}$ of LAM$_{pe}$ 
					& \textbf{1.19\%/0.90\%/0.93} 
					& -/-/- & -/-/- & -/-/- & -/-/- & -/-/- \\
					
					Case 1 & RMSE/MAE/R$^{2}$ of LAM$_{ne}$ 
					& \textbf{1.04\%/0.77\%/0.94} 
					& -/-/- & -/-/- & -/-/- & -/-/- & -/-/- \\
					
					& RMSE/MAE/R$^{2}$ of LLI 
					& \textbf{1.06\%/0.79\%/0.96} 
					& -/-/- & -/-/- & -/-/- & -/-/- & -/-/- \\
					\midrule
					
					& RMSE/MAE/R$^{2}$ of LAM$_{pe}$ 
					& 6.41\%/4.05\%/-0.82 
					& 4.67\%/3.26\%/-0.10 
					& 1.33\%/1.01\%/0.92 
					& 10.39\%/9.38\%/-4.42 
					& 1.19\%/0.89\%/0.94 
					& \textbf{1.20\%/0.91\%/0.93} \\
					
					Case 2 & RMSE/MAE/R$^{2}$ of LAM$_{ne}$ 
					& 7.76\%/4.72\%/-2.19 
					& 6.50\%/4.64\%/-1.55 
					& 2.64\%/1.80\%/0.63 
					& 9.67\%/8.77\%/-4.64 
					& 1.55\%/1.08\%/0.87 
					& \textbf{2.36\%/1.71\%/0.66} \\
					
					& RMSE/MAE/R$^{2}$ of LLI 
					& 7.60\%/4.66\%/-0.91 
					& 6.39\%/4.63\%/-0.51 
					& 2.01\%/1.44\%/0.87 
					& 11.76\%/10.55\%/-4.11 
					& 1.30\%/0.99\%/0.94 
					& \textbf{1.70\%/1.33\%/0.89} \\
					\midrule
					
					& RMSE/MAE/R$^{2}$ of LAM$_{pe}$ 
					& 9.92\%/8.73\%/-3.44 
					& 5.71\%/4.23\%/-0.66 
					& 3.47\%/2.47\%/0.46 
					& 10.14\%/9.12\%/-4.22 
					& 1.52\%/1.15\%/0.90 
					& \textbf{1.86\%/1.45\%/0.82} \\
					
					Case 3 & RMSE/MAE/R$^{2}$ of LAM$_{ne}$ 
					& 9.53\%/8.42\%/-3.57 
					& 5.79\%/4.33\%/-0.90 
					& 3.87\%/2.72\%/0.25 
					& 9.75\%/8.79\%/-4.39 
					& 1.58\%/1.22\%/0.87 
					& \textbf{2.78\%/2.14\%/0.56} \\
					
					& RMSE/MAE/R$^{2}$ of LLI 
					& 11.61\%/10.14\%/-3.23 
					& 6.57\%/4.84\%/-0.50 
					& 5.12\%/3.60\%/0.18 
					& 11.87\%/10.59\%/-3.91 
					& 1.54\%/1.21\%/0.93 
					& \textbf{3.14\%/2.44\%/0.66} \\
					\midrule
					
					& RMSE/MAE/R$^{2}$ of LAM$_{pe}$ 
					& 1.69\%/1.33\%/0.86 
					& 1.84\%/1.44\%/0.81 
					& 1.66\%/1.33\%/0.86 
					& 9.57\%/8.57\%/-4.07 
					& 1.38\%/1.06\%/0.91 
					& \textbf{1.48\%/1.19\%/0.88} \\
					
					Case 4 & RMSE/MAE/R$^{2}$ of LAM$_{ne}$ 
					& 1.84\%/1.43\%/0.84 
					& 2.09\%/1.65\%/0.77 
					& 1.82\%/1.41\%/0.84 
					& 9.55\%/8.52\%/-3.92 
					& 1.39\%/1.08\%/0.91 
					& \textbf{1.80\%/1.40\%/0.83} \\
					
					& RMSE/MAE/R$^{2}$ of LLI 
					& 1.75\%/1.41\%/0.90 
					& 2.03\%/1.61\%/0.86 
					& \textbf{1.65\%/1.33\%/0.91} 
					& 11.41\%/10.08\%/-3.56 
					& 1.36\%/1.07\%/0.94 
					& 1.67\%/1.32\%/0.90 \\
					\midrule
					
					& RMSE/MAE/R$^{2}$ of LAM$_{pe}$ 
					& 2.00\%/1.48\%/0.82 
					& 2.24\%/1.64\%/0.75 
					& 1.77\%/1.37\%/0.86 
					& 10.34\%/9.34\%/-4.43 
					& 1.52\%/1.14\%/0.90 
					& \textbf{1.59\%/1.23\%/0.87} \\
					
					Case 5 & RMSE/MAE/R$^{2}$ of LAM$_{ne}$ 
					& 2.06\%/1.55\%/0.77 
					& 2.22\%/1.69\%/0.70 
					& 1.91\%/1.46\%/0.81 
					& 9.70\%/8.81\%/-4.75 
					& 1.43\%/1.07\%/0.89 
					& \textbf{1.66\%/1.28\%/0.83} \\
					
					& RMSE/MAE/R$^{2}$ of LLI 
					& 2.26\%/1.67\%/0.84 
					& 2.44\%/1.80\%/0.79 
					& 1.89\%/1.46\%/0.89 
					& 12.05\%/10.83\%/-4.19 
					& 1.43\%/1.09\%/0.94 
					& \textbf{1.56\%/1.22\%/0.91} \\
					
					\bottomrule
			\end{tabular}}
		\end{threeparttable}
	\end{table*}
	
	By integrating both the source domain and the few-shot target samples into the generation process, the proposed method shows clear advantages, outperforming Methods 1, 2, and 3. Furthermore, Method 4 is introduced as a theoretical upper bound for the domain generation framework, as it randomly samples the target domain data across the entire lifespan, i.e., data leakage. Despite being restricted to only the first 5\% of early-life target data, the proposed method achieves performance close to this theoretical upper bound, particularly for $\text{LAM}_{pe}$, where its RMSE of 1.20\% nearly matches the 1.19\% obtained by Method 4. The estimation error for $\text{LAM}_{ne}$ remains relatively higher at 2.36\%. This can be attributed to the inherent diagnostic challenge that negative electrode degradation features---particularly in silicon-graphite blended anodes \cite{liu2026advanced}---are often less dominant and rapidly smoothed or masked within complex full-cell dynamic profiles \cite{geslin2025dynamic,wan2026managing}, rendering the negative electrode capacity significantly harder to estimate reliably than the positive electrode \cite{che2026battery}. Notably, Method 4 reaches essentially the same $\text{LAM}_{pe}$ RMSE as the Case 1 benchmark (both 1.19\%), highlighting both the value of full-lifespan target information and the difficulty of achieving comparable performance under a strictly early-life few-shot setting.
	
\begin{figure*}[!h]
		\centering
		\includegraphics[width=16cm]{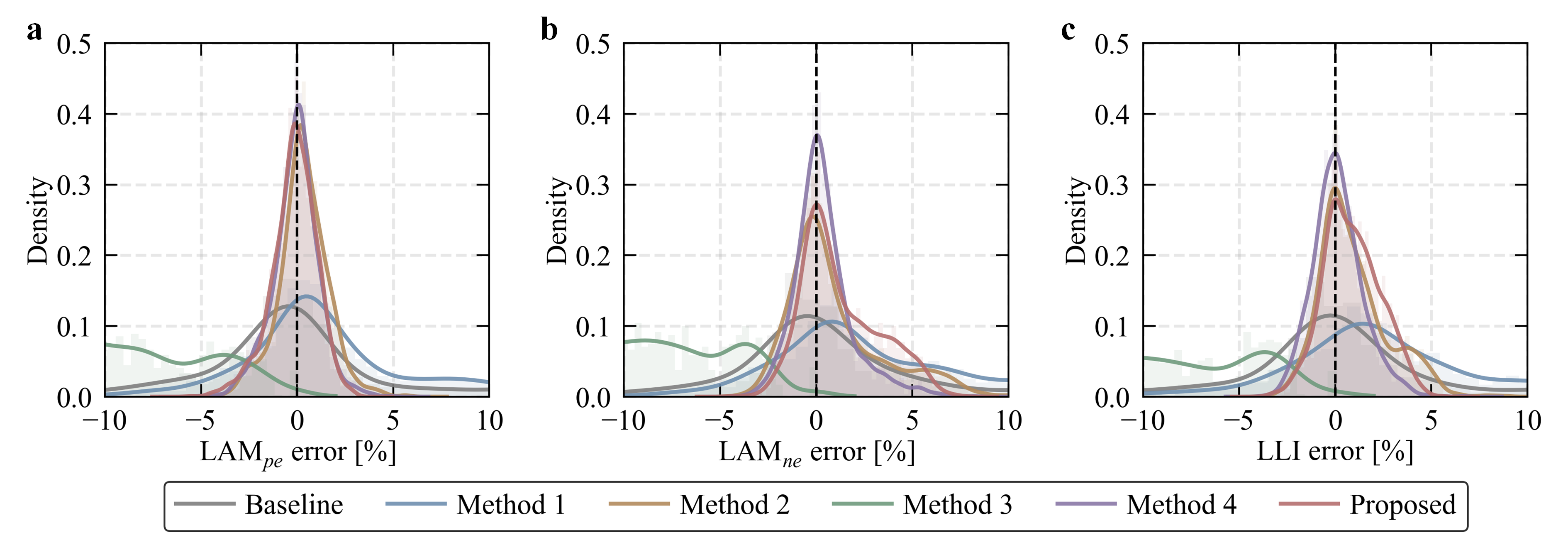}
		\caption{Comparative diagnosis error distribution of different methods under Case 2. (a) for LAM$_{pe}$, (b) for LAM$_{ne}$, (c) for LLI. Source domain: CC discharging samples, Target domain: Synthetic/Drive discharging samples.} 
		\label{fig11_comparison}
	\end{figure*}

	Fig. \ref{fig11_comparison} visually corroborates these quantitative findings by plotting the error density distributions for Case 2. The error curves for the Baseline method, Method 1, and Method 3 are wide and flat, demonstrating poor estimation accuracy. In contrast, Method 2, Method 4, and the proposed method exhibit sharp, zero-centered error distributions. Notably, the error density of the proposed method aligns with the narrow variance observed in Method 4, thereby verifying its effectiveness.
	
	Beyond Case 2, the proposed framework maintains low diagnosis errors across the other scenarios, with detailed results provided in Figs. S5--S13. In the most data-scarce Case 3, where only eight C/16 source cells are available, the proposed method achieves RMSEs of 1.86\%, 2.78\%, and 3.14\% for LAM${pe}$, LAM${ne}$, and LLI, respectively. The corresponding RMSEs decrease to 1.48\%, 1.80\%, and 1.67\% in Case 4 and 1.59\%, 1.66\%, and 1.56\% in Case 5. An exception occurs for LLI in Case 4, where Method 2 yields a slightly lower RMSE than the proposed method (1.65\% versus 1.67\%). The associated error distributions and detailed prediction results are provided in Figs. S6--S7, S9--S10, and S12--S13, respectively.
	
	The Bayesian optimization results further reveal that the preferred generation strategy depends strongly on source domain richness (Figs. S5, S8, and S11). The generation multiplier $M_{gen}$ dominates in the highly data-scarce Case 3, whereas Case 4 exhibits a more balanced sensitivity across multiple hyperparameters. In the data-richer Case 5, the target domain sampling probability $P_{sample}$ becomes the dominant factor. These contrasting patterns indicate that generation volume is particularly important when source information is scarce, while source-target balancing becomes increasingly important as the source representation becomes richer. This scenario-dependent behavior supports automated optimization rather than prescribing a fixed generation strategy across different domains.
	
	\section{Conclusion \label{conclusion}}
	
	This study establishes a unified closed-loop domain generation framework for cross-domain battery health diagnosis by addressing the severe shifts in feature distributions and inherent data scarcity. Within this framework, physics-constrained domain generation is implemented through VAE-based latent extrapolation coupled with automated Bayesian optimization. The VAE expands target-oriented latent features, while Bayesian optimization calibrates sample mixing, latent regularization, extrapolation, perturbation, and generation volume, thereby generating data within physically feasible degradation boundaries. Moreover, by extracting diagnostic features from partial dynamic discharging data, the framework enables onboard battery health diagnosis of LAM$_{pe}$, LAM$_{ne}$, and LLI without requiring full-lifespan target domain experiments.
	
	Extensive evaluations across four cross-domain scenarios explicitly validate the superiority of the proposed method. In the most typical Lab2Real scenario, the proposed method achieves RMSE values to 1.20\%, 2.36\%, and 1.70\%, respectively. Furthermore, ablation studies confirmed that the early-life few-shot domain generation efficiently reconstructs the missing target domain. The physical consistency of the generated domain is further demonstrated by the accurate reconstruction of differential voltage curves and quantitative analysis that capture the evolution of phase-transition peaks over the battery lifespan.
	
	From a broader perspective, this research provides a new perspective on battery health diagnosis. While existing literature predominantly focuses on designing increasingly complex diagnostic network architectures or hunting for novel dynamic features, this study highlights that enhancing the quality of the underlying data distribution through physics-constrained domain generation is an effective yet largely untapped pathway. Since the proposed domain generation module is inherently decoupled from the downstream diagnosis networks, it acts as a unified, plug-and-play data enhancer that can be seamlessly integrated with any existing data-driven battery management system to boost cross-domain robustness.

	\section*{Appendix}
	
	For different cross-domain cases (Case 2/3/4/5), the proposed method achieves average RMSE values of 1.53\%, 2.15\%, and 2.02\% for the loss of active material in the positive electrode, loss of active material in the negative electrode, and loss of lithium inventory, respectively. Detailed results of Cases 1--5 are listed in Table \ref{tab2}, and the corresponding figures and VAE/MLP network structures are provided in the Supplementary Materials.
	
	\bibliographystyle{elsarticle-num}
	\bibliography{Reference}
	
\end{document}